\documentclass[aps,prl,reprint,longbibliography,superscriptaddress]{revtex4-1}
\usepackage{mathrsfs}
\usepackage{amsmath,gensymb}
\usepackage{amsfonts}
\usepackage{amssymb}
\usepackage{amsthm}
\usepackage{graphicx}
\usepackage{natbib}
\usepackage{color}
\usepackage{hyperref}
\usepackage{bm}
\usepackage[caption=false]{subfig}
\usepackage{verbatim}

\begin{document}
\title{Anisotropic and controllable Gilbert-Bloch dissipation in spin valves}

\author{Akashdeep Kamra}
\email{akashdeep.kamra@ntnu.no}
\affiliation{Center for Quantum Spintronics, Department of Physics, Norwegian University of Science and Technology, Trondheim, Norway}

\author{Dmytro M. Polishchuk}
\affiliation{Nanostructure Physics, Royal Institute of Technology, Stockholm, Sweden}

\author{Vladislav Korenivski}
\affiliation{Nanostructure Physics, Royal Institute of Technology, Stockholm, Sweden}

\author{Arne Brataas}
\affiliation{Center for Quantum Spintronics, Department of Physics, Norwegian University of Science and Technology, Trondheim, Norway}

\begin{abstract}
Spin valves form a key building block in a wide range of spintronic concepts and devices from magnetoresistive read heads to spin-transfer-torque oscillators. We elucidate the dependence of the magnetic damping in the free layer on the angle its equilibrium magnetization makes with that in the fixed layer. The spin pumping-mediated damping is anisotropic and tensorial, with Gilbert- and Bloch-like terms. Our investigation reveals a mechanism for tuning the free layer damping in-situ from negligible to a large value via the orientation of fixed layer magnetization, especially when the magnets are electrically insulating. Furthermore, we expect the Bloch contribution that emerges from the longitudinal spin accumulation in the non-magnetic spacer to play an important role in a wide range of other phenomena in spin valves.
\end{abstract}

\maketitle

{\it Introduction.} -- The phenomenon of magnetoresistance is at the heart of contemporary data storage technologies~\cite{Fert2008,Parkin2003}. The dependence of the resistance of a multilayered heterostructure comprising two or more magnets on the angles between their respective magnetizations has been exploited to read magnetic bits with a high spatial resolution~\cite{Nagasaka2009}. Furthermore, spin valves comprised of two magnetic layers separated by a non-magnetic conductor have been exploited in magnetoresistive random access memories~\cite{Akerman2005,Parkin2003,Bhatti2017}. Typically, in such structures, one `free layer' is much thinner than the other `fixed layer' allowing for magnetization dynamics and switching in the former. The latter serves to spin-polarize the charge currents flowing across the device and thus exert spin-torques on the former~\cite{Brataas2012b,Berger1996,Slonczewski1996,Ralph2008}. Such structures exhibit a wide range of phenomena from magnetic switching~\cite{Bhatti2017} to oscillations~\cite{Kim2012,Silva2008} driven by applied electrical currents.

With the rapid progress in taming pure spin currents~\cite{Bauer2012,Kruglyak2010,Uchida2010,Adachi2013,Chumak2015,Maekawa2012,Hirsch1999,Saitoh2006,Weiler2013}, magnetoresistive phenomena have found a new platform in hybrids involving magnetic insulators (MIs). The electrical resistance of a non-magnetic metal (N) was found to depend upon the magnetic configuration of an adjacent insulating magnet~\cite{Nakayama2013,Huang2012,Althammer2013,Vlietstra2013}. This phenomenon, dubbed spin Hall magnetoresistance (SMR), relies on the pure spin current generated via spin Hall effect (SHE) in N~\cite{Chen2013,Chen2016}. The SHE spin current accumulates spin at the MI/N interface, which is absorbed by the MI depending on the angle between its magnetization and the accumulated spin polarization. The net spin current absorbed by the MI manifests as additional magnetization-dependent contribution to resistance in N via the inverse SHE. The same principle of magnetization-dependent spin absorption by MI has also been exploited in demonstrating spin Nernst effect~\cite{Meyer2017}, i.e. thermally generated pure spin current, in platinum.  

While the ideas presented above have largely been exploited in sensing magnetic fields and magnetizations, tunability of the system dissipation is a valuable, underexploited consequence of magnetoresistance. Such an electrically controllable resistance of a magnetic wire hosting a domain wall~\cite{Parkin2008} has been suggested as a basic circuit element~\cite{Wang2009} in a neuromorphic computing~\cite{Mead1990} architecture. In addition to the electrical resistance or dissipation, the spin valves should allow for controlling the magnetic damping in the constituent magnets~\cite{Tserkovnyak2005}. Such an in-situ control can be valuable in, for example, architectures where a magnet is desired to have a large damping to attain low switching times and a low dissipation for spin dynamics and transport~\cite{Kruglyak2010,Chumak2015}. Furthermore, a detailed understanding of magnetic damping in spin valves is crucial for their operation as spin-transfer-torque oscillators~\cite{Kim2012} and memory cells~\cite{Bhatti2017}. 

\begin{figure}[tb]
	\begin{center}
		\includegraphics[width=55mm]{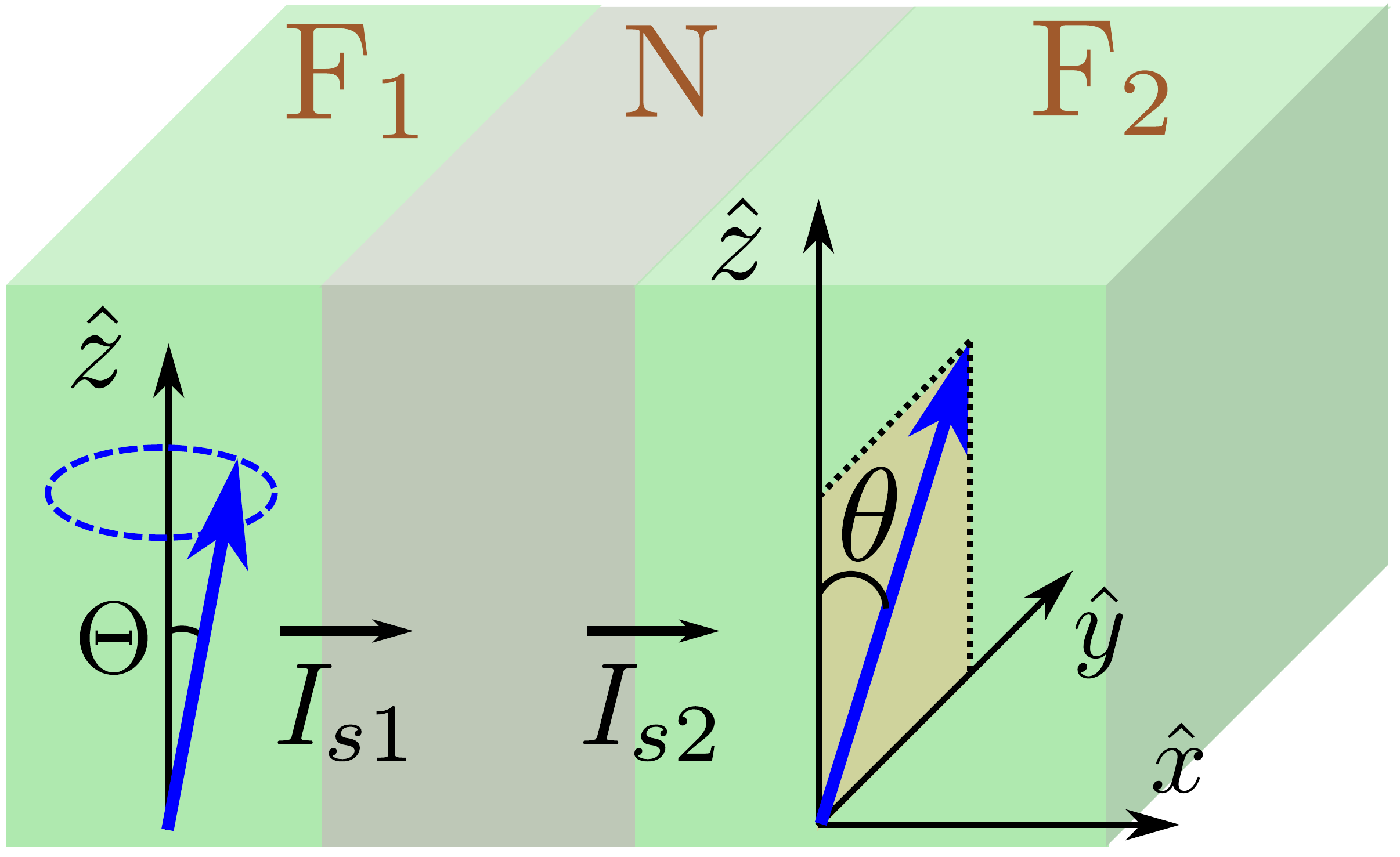}
		\caption{Schematic depiction of the device under investigation. The blue arrows denote the magnetizations. The fixed layer $\mathrm{F}_2$ magnetization remains static. The free layer $\mathrm{F}_1$ magnetization precesses about the z-axis with an average cone angle $\Theta \ll 1$. The two layers interact dynamically via spin pumping and backflow currents.}
		\label{fig:schematic}
	\end{center}
\end{figure}

Inspired by these new discoveries~\cite{Nakayama2013,Meyer2017} and previous related ideas~\cite{Tserkovnyak2005,Tserkovnyak2003,Taniguchi2007,Polishcuk2018}, we suggest new ways of tuning the magnetic damping of the free layer $\mathrm{F}_1$ in a spin valve (Fig. \ref{fig:schematic}) via controllable absorption by the fixed layer $\mathrm{F}_2$ of the spin accumulated in the spacer N due to spin pumping~\cite{Tserkovnyak2002,Tserkovnyak2005}. The principle for this control over spin absorption is akin to the SMR effect discussed above and capitalizes on altering the $\mathrm{F}_2$ magnetization direction. When spin relaxation in N is negligible, the spin lost by $\mathrm{F}_1$ is equal to the spin absorbed by $\mathrm{F}_2$. This lost spin appears as tensorial Gilbert~\cite{Gilbert2004} and Bloch~\cite{Bloch1946} damping in $\mathrm{F}_1$ magnetization dynamics. In its isotropic form, the Gilbert contribution arises due to spin pumping and is well established~\cite{Berger2001,Tserkovnyak2002,Tserkovnyak2003,Tserkovnyak2005,Mosendz2008,Taniguchi2007,Chiba2015}. We reveal that the Bloch term results from backflow due to a finite dc longitudinal spin accumulation in N. Our results for the angular and tensorial dependence of the Gilbert damping are also, to best of our knowledge, new. 

\begin{figure}[tb]
	\begin{center}
		\includegraphics[width=65mm]{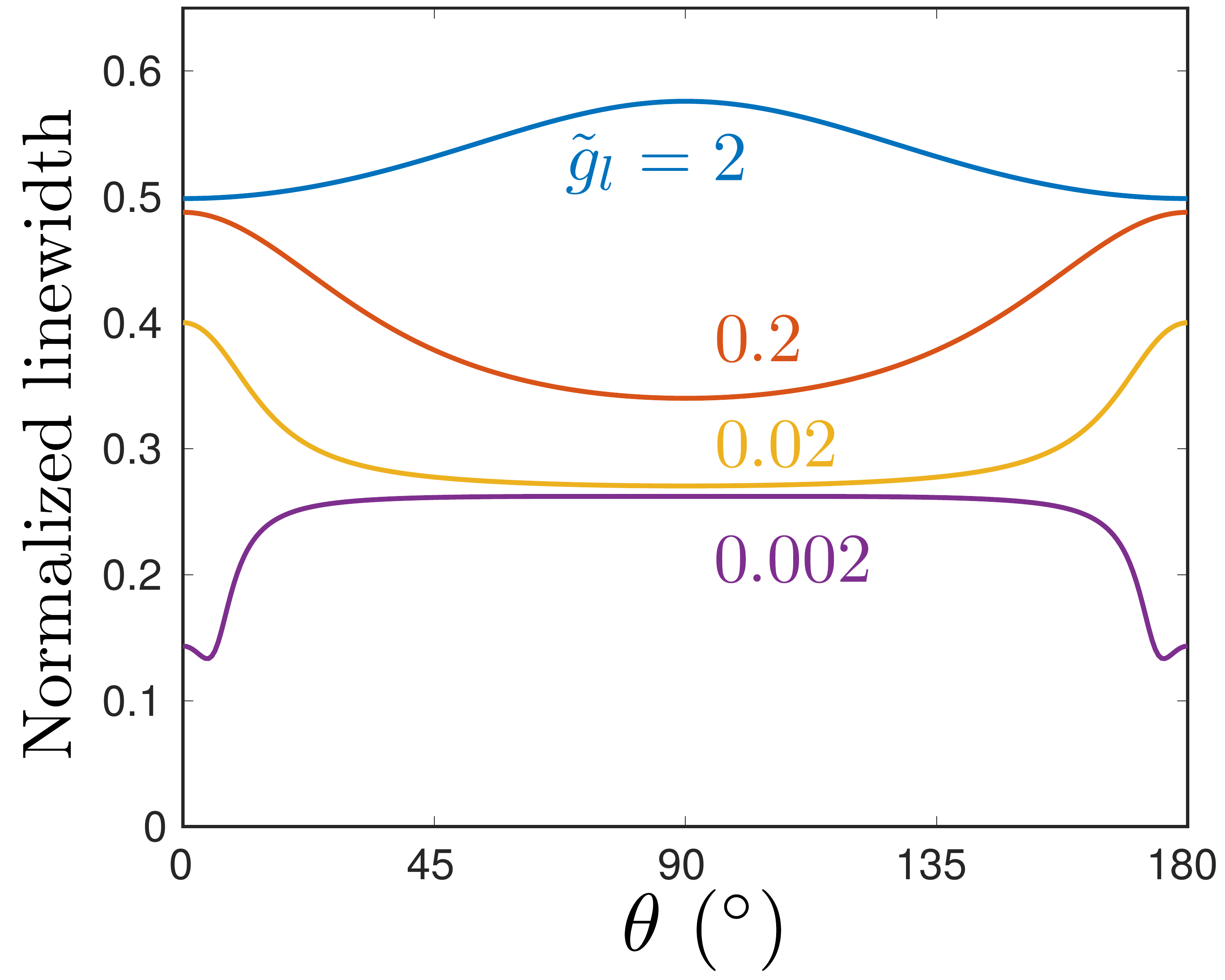}
		\caption{Normalized ferromagnetic resonance (FMR) linewidth of $\mathrm{F}_1$ for different values of the longitudinal spin conductance $\tilde{g}_l \equiv g_l/g_r$ of N$|\mathrm{F}_2$ bilayer. The various parameters employed are $\tilde{g}_r^\prime \equiv g_r^\prime/g_r = 1$, $\Theta = 0.1$ rad, $\omega_0 = 10 \times 2 \pi$ GHz, and $\omega_{ax} = 1 \times 2 \pi$ GHz. $g_r$ and $g_r^
		\prime$ are the spin-mixing conductances of $\mathrm{F}_1|$N and N$|\mathrm{F}_2$ interfaces respectively. Only the spin pumping-mediated contribution to the linewidth has been considered and is normalized to its value for the case of spin pumping into a perfect spin sink~\cite{Tserkovnyak2005}.}
		\label{fig:gldep}
	\end{center}
\end{figure}

We show that the dissipation in $\mathrm{F}_1$, expressed in terms of ferromagnetic resonance (FMR) linewidth, varies with the angle $\theta$ between the two magnetizations (Fig. \ref{fig:gldep}). The maximum dissipation is achieved in collinear or orthogonal configurations depending on the relative size of the spin-mixing $g_{r}^\prime$ and longitudinal spin $g_l$ conductances of the N$|\mathrm{F}_2$ subsystem. For very low $g_l$, which can be achieved employing insulating magnets, the spin pumping mediated contribution to the linewidth vanishes for collinear configurations and attains a $\theta$-independent value for a small non-collinearity. This can be used to strongly modulate the magnetic dissipation in $\mathrm{F}_1$ electrically via, for example, an $\mathrm{F}_2$ comprised by a magnetoelectric material~\cite{Fusil2014}.

{\it FMR linewidth.} -- Disregarding intrinsic damping for convenience, the magnetization dynamics of $\mathrm{F}_1$ including a dissipative spin transfer torque arising from the spin current lost $\pmb{I}_{s1}$ may be expressed as: 
\begin{align}\label{eq:stt}
\dot{\hat{\pmb{m}}} = & - |\gamma| \left( \hat{\pmb{m}} \times \mu_0 \pmb{H}_{\mathrm{eff}} \right) + \frac{|\gamma|}{M_s V} \pmb{I}_{s1}.
\end{align}
Here, $\hat{\pmb{m}}$ is the unit vector along the $\mathrm{F}_1$ magnetization $\pmb{M}$ treated within the macrospin approximation, $\gamma~(< 0)$ is the gyromagnetic ratio, $M_s$ is the saturation magnetization, $V$ is the volume of $\mathrm{F}_1$, and $\pmb{H}_{\mathrm{eff}}$ is the effective magnetic field. Under certain assumptions of linearity as will be detailed later, Eq, (\ref{eq:stt}) reduces to the Landau-Lifshitz equation with Gilbert-Bloch damping~\cite{Gilbert2004,Bloch1946}:
\begin{align}\label{eq:llgb}
\dot{\hat{\pmb{m}}} = & - |\gamma| \left( \hat{\pmb{m}} \times \mu_0 \pmb{H}_{\mathrm{eff}} \right) + \left( \hat{\pmb{m}} \times \pmb{G} \right) - \pmb{B}.
\end{align}
Considering the equilibrium orientation $\hat{\pmb{m}}_{\mathrm{eq}} = \hat{\pmb{z}}$, Eq. (\ref{eq:llgb}) is restricted to the small transverse dynamics described by $m_{x,y} \ll 1$, while the z-component is fully determined by the constraint $\hat{\pmb{m}} \cdot \hat{\pmb{m}} = 1$. Parameterized by a diagonal dimensionless tensor $\check{\alpha}$, the Gilbert damping has been incorporated via $\pmb{G} = \alpha_{xx} \dot{m}_x \hat{\pmb{x}} + \alpha_{yy} \dot{m}_y \hat{\pmb{y}}$ in Eq. (\ref{eq:llgb}). The Bloch damping is parametrized via a diagonal frequency tensor $\check{\Omega}$ as $\pmb{B} = \Omega_{xx} m_x \hat{\pmb{x}} + \Omega_{yy} m_y \hat{\pmb{y}}$. A more familiar, although insufficient for the present considerations, form of Bloch damping can be obtained by assuming isotropy in the transverse plane: $\pmb{B} = \Omega_0 \left( \hat{\pmb{m}} - \hat{\pmb{m}}_{\mathrm{eq}} \right)$. This form, restricted to transverse dynamics, makes its effect as a relaxation mechanism with characteristic time $1/\Omega_0$ evident. The Bloch damping, in general, captures the so-called inhomogeneous broadening and other, frequency independent contributions to the magnetic damping.  

Considering uniaxial easy-axis and easy-plane anisotropies, parametrized respectively by $K_z$ and $K_x$~\footnote{The easy-plane may stem from the shape anisotropy in thin films, in which case $K_x = \mu_0/2$ while the easy-axis may be magnetocrystalline in nature~\cite{Akhiezer1968}.}, the magnetic free energy density $F_m$ is expressed as: $F_m =  - \mu_0 \pmb{M}\cdot\pmb{H}_{\mathrm{ext}} - K_z M_{z}^2 + K_{x} M_{x}^2,$ with $\pmb{H}_{\mathrm{ext}} = H_0 \hat{\pmb{z}} + \pmb{h}_{rf}$ as the applied static plus microwave field. Employing the effective field $\mu_0 \pmb{H}_{\mathrm{eff}} = - \partial F_m/\partial \pmb{M}$ in Eq. (\ref{eq:llgb}) and switching to Fourier space [$\sim \exp(i \omega t)$], we obtain the resonance frequency $\omega_r = \sqrt{\omega_0 (\omega_0 + \omega_{ax})}$. Here, $\omega_0 \equiv |\gamma| (\mu_0 H_0 + 2 K_z M_s)$ and $\omega_{ax} \equiv |\gamma| 2 K_x M_s$. The FMR linewidth is evaluated as:
\begin{align}\label{eq:linewidth}
|\gamma| \mu_0 \Delta H  = & \frac{\left(\alpha_{xx} + \alpha_{yy}\right)}{2} \omega + t~ \frac{\left(\Omega_{xx} + \Omega_{yy} \right)}{2} \nonumber \\
  & + \frac{t \omega_{ax}}{4} \left( \alpha_{yy} - \alpha_{xx} \right),
\end{align} 
where $\omega$ is the frequency of the applied microwave field $\pmb{h}_{rf}$ and is approximately $\omega_r$ close to resonance, and $t \equiv \omega / \sqrt{\omega^2 + \omega_{ax}^2 /4} \approx 1$ for a weak easy-plane anisotropy. Thus, in addition to the anisotropic Gilbert contributions, the Bloch damping provides a nearly frequency-independent offset in the linewidth.

\begin{figure}[tb]
	\begin{center}
		\includegraphics[width=65mm]{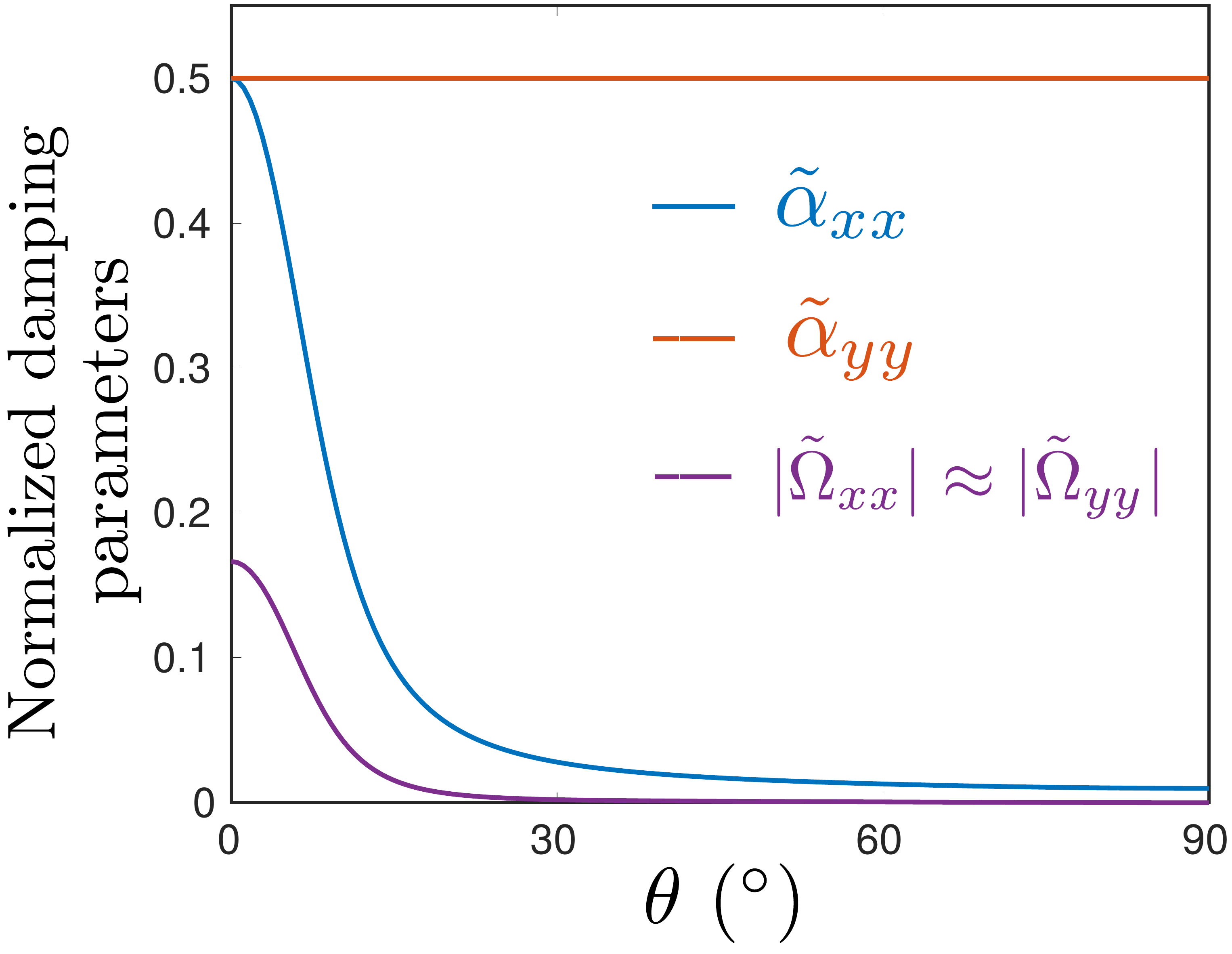}
		\caption{Normalized damping parameters for $\mathrm{F}_1$ magnetization dynamics vs. spin valve configuration angle $\theta$ (Fig. \ref{fig:schematic}). $\tilde{\alpha}_{xx} \neq \tilde{\alpha}_{yy}$ signifies the tensorial nature of the Gilbert damping. The Bloch parameters $\tilde{\Omega}_{xx} \approx \tilde{\Omega}_{yy}$ are largest for the collinear configuration. The curves are mirror symmetric about $\theta = 90\degree$. $\tilde{g}_r^\prime = 1$, $\tilde{g}_l = 0.01$, $\Theta = 0.1$, $\omega_0 = 10 \times 2 \pi$ GHz, and $\omega_{ax} = 1 \times 2 \pi$ GHz.}
		\label{fig:damppar}
	\end{center}
\end{figure}

{\it Spin flow.} -- We now examine spin transport in the device with the aim of obtaining the damping parameters that determine the linewidth [Eq. (\ref{eq:linewidth})]. The N layer is considered thick enough to eliminate static exchange interaction between the two magnetic layers~\cite{Tserkovnyak2005,Chiba2015}. Furthermore, we neglect the imaginary part of the spin-mixing conductance, which is small in metallic systems and does not affect dissipation in any case. Disregarding longitudinal spin transport and relaxation in the thin free layer, the net spin current $\pmb{I}_{s1}$ lost by $\mathrm{F}_1$ is the difference between the spin pumping and backflow currents~\cite{Tserkovnyak2005}:
\begin{align}\label{eq:is1}
\pmb{I}_{s1} = & \frac{g_r}{4 \pi} \left( \hbar ~ \hat{\pmb{m}} \times \dot{\hat{\pmb{m}}} - \hat{\pmb{m}} \times \pmb{\mu}_s \times \hat{\pmb{m}} \right),
\end{align}
where $g_r$ is the real part of the $\mathrm{F}_1|$N interfacial spin-mixing conductance, and $\pmb{\mu}_s$ is the spatially homogeneous spin accumulation in the thin N layer. The spin current absorbed by $\mathrm{F}_2$ may be expressed as~\cite{Tserkovnyak2005}:
\begin{align}\label{eq:is2}
\pmb{I}_{s2} & =  \frac{g_r^\prime}{4 \pi} \hat{\pmb{m}}_2 \times \pmb{\mu}_s \times \hat{\pmb{m}}_2 + \frac{g_l}{4 \pi} \left( \hat{\pmb{m}}_2 \cdot \pmb{\mu}_s  \right) \hat{\pmb{m}}_2, \nonumber \\
   & \equiv \sum_{i,j = \left\{x,y,z \right\} } \frac{g_{ij}}{4 \pi} \mu_{sj} \hat{\pmb{i}},
\end{align}
where $g_l$ and $g_r^\prime$ are respectively the longitudinal spin conductance and the real part of the interfacial spin-mixing conductance of the N$|\mathrm{F}_2$ subsystem, $\hat{\pmb{m}}_2$ denotes the unit vector along $\mathrm{F}_2$ magnetization, and $g_{ij} = g_{ji}$ are the components of the resulting total spin conductance tensor. $g_l$ quantifies the absorption of the spin current along the direction of $\hat{\pmb{m}}_2$, the so-called longitudinal spin current. For metallic magnets, it is dominated by the diffusive spin current carried by the itinerant electrons, which is dissipated over the spin relaxation length~\cite{Tserkovnyak2005}. On the other hand, for insulating magnets, the longitudinal spin absorption is dominated by magnons~\cite{Cornelissen2015,Goennenwein2015} and is typically much smaller than for the metallic case, especially at low temperatures.  Considering $\hat{\pmb{m}}_2 = \sin \theta ~ \hat{\pmb{y}} + \cos \theta ~ \hat{\pmb{z}}$ (Fig. \ref{fig:schematic}), Eq. (\ref{eq:is2}) yields $g_{xx} = g_r^\prime$, $g_{yy} = g_r^\prime \cos^2 \theta + g_l \sin^2 \theta$, $g_{zz} = g_r^\prime \sin^2 \theta + g_l \cos^2 \theta$,
$g_{xy} = g_{yx} = g_{xz} = g_{zx} = 0$, and $g_{yz} = g_{zy} = (g_l - g_r^\prime) \sin \theta \cos \theta $.

Relegating the consideration of a small but finite spin relaxation in the thin N layer to the supplemental material~\cite{SupplMat}, we assume here that the spin current lost by $\mathrm{F}_1$ is absorbed by $\mathrm{F}_2$, i.e., $\pmb{I}_{s1} = \pmb{I}_{s2}$. Imposing this spin current conservation condition, the spin accumulation in N along with the currents themselves can be determined. We are primarily interested in the transverse (x and y) components of the spin current since these fully determine the magnetization dynamics ($\hat{\pmb{m}} \cdot \hat{\pmb{m}} = 1$):
\begin{equation}\label{eq:is}
\begin{aligned}
I_{s1x} = & \frac{1}{4 \pi} \frac{g_r g_{xx}}{g_r + g_{xx}} \left( - \hbar \dot{m}_y + m_x \mu_{sz} \right), \\
I_{s1y} = & \frac{1}{4 \pi} \left[ \frac{g_r g_{yy}}{g_r + g_{yy}} \left( \hbar \dot{m}_x + m_y \mu_{sz} \right) + g_{yz} \mu_{sz} (1 - l_y) \right], \\
\mu_{sz} = & \frac{\hbar g_r \left( l_x m_x \dot{m}_y - l_y m_y \dot{m}_x - p \dot{m}_x \right)}{g_{zz} - p g_{yz} + g_r \left( l_x m_x^2 + l_y m_y^2 + 2 p m_y\right)},
\end{aligned}
\end{equation}
where $l_{x,y} \equiv g_{xx,yy}/(g_r + g_{xx,yy})$ and $p \equiv g_{yz}/(g_r + g_{yy})$. The spin lost by $\mathrm{F}_1$ appears as damping in the magnetization dynamics [Eqs. (\ref{eq:stt}) and (\ref{eq:llgb})]~\cite{Tserkovnyak2002,Tserkovnyak2005}.

We pause to comment on the behavior of $\mu_{sz}$ thus obtained [Eq. (\ref{eq:is})]. Typically, $\mu_{sz}$ is considered to be first or second order in the cone angle, and thus negligibly small. However, as discussed below, an essential new finding is that it becomes independent of the cone angle and large under certain conditions. For a collinear configuration and vanishing $g_l$, $g_{zz} = g_{yz} = 0$ results in $\tilde{\mu}_{sz} \equiv \mu_{sz}/ \hbar \omega \to 1$ ~\cite{Berger2001}. Its finite dc value contributes to the Bloch damping [Eq. (\ref{eq:is})]~\cite{Berger2001}. For a non-collinear configuration, $\mu_{sz} \approx - \hbar g_r p \dot{m}_x / (g_{zz} - p g_{yz})$ and contributes to Gilbert damping via $I_{s1y}$ [Eq. (\ref{eq:is})]. Thus, in general, we may express the spin accumulation as $\mu_{sz} = \mu_{sz0} + \mu_{sz1}$~\footnote{As detailed in the supplemental material~\cite{SupplMat}, we have disregarded the term in $\mu_{sz}$ which oscillates with a frequency $2 \omega$. Strictly speaking, this term needs to be included even in our linear analysis, since it produces terms oscillating with $\omega$ when multiplied with another term at $\omega$. However, such a contribution is only relevant in a narrow parameter range which may be hard to resolve in an experiment. Furthermore, it requires a non-linear solution to the equations and is beyond the scope of the present work.}, where $\mu_{sz0}$ is the dc value and $\mu_{sz1} \propto \dot{m}_{x}$ is the linear oscillating component. $\mu_{sz0}$ and $\mu_{sz1}$ contribute, respectively, to Bloch and Gilbert damping. 


{\it Gilbert-Bloch dissipation.} -- Equations (\ref{eq:stt}) and (\ref{eq:is}) completely determine the magnetic damping in $\mathrm{F}_1$. However, these equations are non-linear and cannot be captured within our linearized framework [Eqs. (\ref{eq:llgb}) and (\ref{eq:linewidth})]. The leading order effects, however, are linear in all but a narrow range of parameters. Evaluating these leading order terms within reasonable approximations detailed in the supplemental material~\cite{SupplMat}, we are able to obtain the Gilbert and Bloch damping tensors $\check{\alpha}$ and $\check{\Omega}$. Obtaining the general result numerically~\cite{SupplMat}, we present the analytic expressions for two cases covering a large range of the parameter space below.

First, we consider the collinear configurations in the limit of $\tilde{g}_l \equiv g_{l}/g_{r} \to 0$. As discussed above, we obtain $\tilde{\mu}_{sz0} \equiv \mu_{sz0}/\hbar \omega \to 1$ and $\tilde{\mu}_{sz1} \equiv \mu_{sz1} / \hbar \omega \to 0$ [Eq. (\ref{eq:is})]. Thus the components of the damping tensors can be directly read from Eq. (\ref{eq:is}) as $\tilde{\alpha}_{xx,yy} \equiv \alpha_{xx,yy}/\alpha_{ss} = l_{y,x} = g_{r}^\prime / (g_r + g_{r}^\prime) = \tilde{g}_{r}^\prime / (1 + \tilde{g}_{r}^\prime),$ and $\tilde{\Omega}_{xx,yy} \equiv \Omega_{xx,yy}/(\alpha_{ss} \omega) = - l_{x,y} \mu_{sz0} / (\hbar \omega) = - g_{r}^\prime / (g_r + g_{r}^\prime) = - \tilde{g}_{r}^\prime / (1 + \tilde{g}_{r}^\prime)$. Here, we defined $\tilde{g}_r^\prime \equiv g_r^\prime / g_r$ and $\alpha_{ss} \equiv \hbar g_r |\gamma|/(4 \pi M_s V)$ is the Gilbert constant for the case of spin-pumping into an ideal spin sink~\cite{Tserkovnyak2002,Tserkovnyak2005}. Substituting these values in Eq. (\ref{eq:linewidth}), we find that the linewidth, or equivalently damping, vanishes. This is understandable since the system we have considered is not able to relax the z component of the spin at all. There can, thus, be no net contribution to magnetic damping. $\mu_{sz0}$ accumulated in N opposes the Gilbert relaxation via a negative Bloch contribution~\cite{Berger2001}. The latter may also be understood as an anti-damping spin transfer torque due to the accumulated spin~\cite{Brataas2012b}.

\begin{figure}[tb]
	\begin{center}
		\includegraphics[width=65mm]{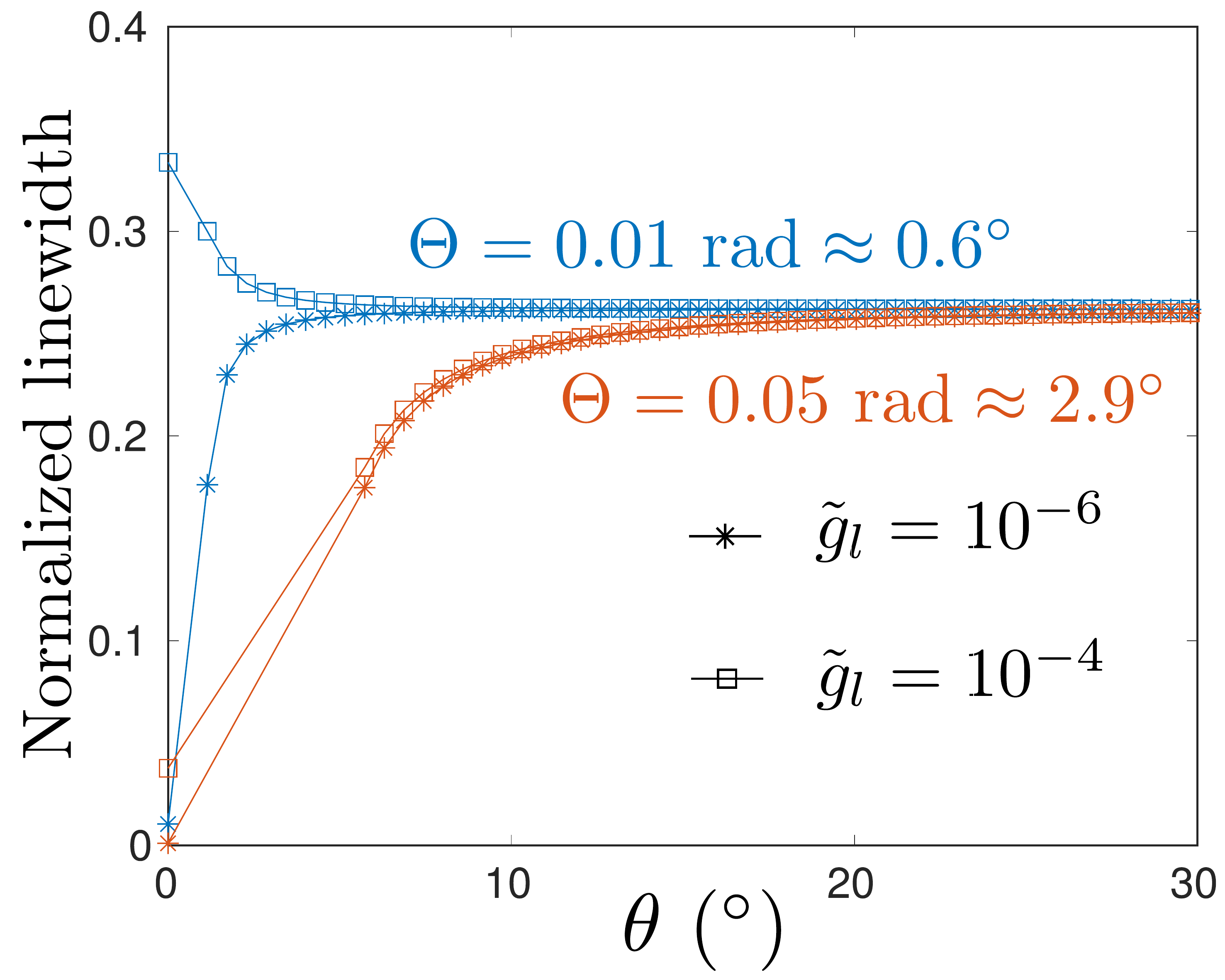}
		\caption{Normalized FMR linewidth of $\mathrm{F}_1$ for very small $\tilde{g}_l$. The squares and circles denote the evaluated points while the lines are guides to the eye. The linewidth increases from being negligible to its saturation value as $\theta$ becomes comparable to the average cone angle $\Theta$. $\tilde{g}_r^\prime = 1$, $\omega_0 = 10 \times 2 \pi$ GHz, and $\omega_{ax} = 1 \times 2 \pi$ GHz.}
		\label{fig:lowgl}
	\end{center}
\end{figure}

Next, we assume the system to be in a non-collinear configuration such that $\tilde{\mu}_{sz0} \to 0$ and may be disregarded, while $\tilde{\mu}_{sz1}$ simplifies to:
\begin{align}
\tilde{\mu}_{sz1} = & -  \frac{\dot{m}_x}{\omega} ~ \frac{(\tilde{g}_l - \tilde{g}_r^\prime) \sin \theta \cos \theta}{\tilde{g}_r^\prime \tilde{g}_l + \tilde{g}_l \cos^2 \theta + \tilde{g}_r^\prime \sin^2 \theta},
\end{align}
where $\tilde{g}_{l} \equiv g_l/g_r$ and $\tilde{g}_r^\prime \equiv g_r^\prime/g_r$ as above. This in turn yields the following Gilbert parameters via Eq. (\ref{eq:is}), with the Bloch tensor vanishing on account of $\tilde{\mu}_{sz0} \to 0$:
\begin{align}\label{eq:gilb}
\tilde{\alpha}_{xx} = & \frac{\tilde{g}_r^\prime \tilde{g}_l}{\tilde{g}_r^\prime \tilde{g}_l + \tilde{g}_l \cos^2 \theta + \tilde{g}_r^\prime \sin^2 \theta}, \quad \tilde{\alpha}_{yy} = \frac{\tilde{g}_r^\prime}{1 + \tilde{g}_r^\prime},
\end{align}
where $\tilde{\alpha}_{xx,yy} \equiv \alpha_{xx,yy}/\alpha_{ss}$ as above. Thus, $\tilde{\alpha}_{yy}$ is $\theta$-independent since $\hat{\pmb{m}}_2$ lies in the y-z plane and the x-component of spin, the absorption of which is captured by $\tilde{\alpha}_{yy}$, is always orthogonal to $\hat{\pmb{m}}_2$. $\tilde{\alpha}_{xx}$, on the other hand, strongly varies with $\theta$ and is generally not equal to $\tilde{\alpha}_{yy}$ highlighting the tensorial nature of the Gilbert damping.

Figure \ref{fig:damppar} depicts the configurational dependence of normalized damping parameters. The Bloch parameters are appreciable only close to the collinear configurations on account of their proportionality to $\mu_{sz0}$. The $\theta$ range over which they decrease to zero is proportional to the cone angle $\Theta$ [Eq. (\ref{eq:is})]. The Gilbert parameters are described sufficiently accurately by Eq. (\ref{eq:gilb}). The linewidth [Eq. (\ref{eq:linewidth})] normalized to its value for the case of spin pumping into a perfect spin sink has been plotted in Fig. \ref{fig:gldep}. For low $\tilde{g}_l$, the Bloch contribution partially cancels the Gilbert dissipation, which results in a smaller linewidth close to the collinear configurations~\cite{Berger2001}. As $\tilde{g}_l$ increases, the relevance of Bloch contribution and $\mu_{sz0}$ diminishes, and the results approach the limiting condition described analytically by Eq. (\ref{eq:gilb}). In this regime, the linewidth dependence exhibits a maximum for either collinear or orthogonal configuration depending on whether $\tilde{g}_l / \tilde{g}_r^\prime$ is smaller or larger than unity. Physically, this change in the angle with maximum linewidth is understood to reflect whether transverse or longitudinal spin absorption is stronger. 

We focus now on the case of very low $\tilde{g}_l$ which can be realized in structures with electrically-insulating magnets. Figure \ref{fig:lowgl} depicts the linewidth dependence close to the collinear configurations. The evaluated points are marked with stars and squares while the lines smoothly connect the calculated points. The gap in data for very small angles reflects the limited validity of our linear theory, as discussed in the supplemental material~\cite{SupplMat}. As per the limiting case $\tilde{g}_l \to 0$ discussed above, the linewidth should vanish in perfectly collinear states. A more precise statement for the validity of this limit is reflected in Fig. \ref{fig:lowgl} and Eq. (\ref{eq:is}) as $\tilde{g}_l/\Theta^2 \to 0$. For sufficiently low $\tilde{g}_l$, the linewidth changes sharply from a negligible value to a large value over a $\theta$ range approximately equal to the cone angle $\Theta$. This shows that systems comprised of magnetic insulators bearing a very low $\tilde{g}_l$ are highly tunable as regards magnetic/spin damping by relatively small deviation from the collinear configuration. The latter may be accomplished electrically by employing magnetoelectric material~\cite{Fusil2014} for $\mathrm{F}_2$ or via current driven spin transfer torques~\cite{Sinova2015,Ralph2008,Brataas2012b}.  

{\it Discussion.} -- Our identification of damping contributions as Gilbert-like and Bloch-like [Eq. (\ref{eq:is})] treats $\mu_{sz}$ as an independent variable that may result from SHE, for example. When it is caused by spin pumping current and $\mu_{sz} \propto \omega$, this Gilbert-Bloch distinction is less clear and becomes a matter of preference. Our results demonstrate the possibility of tuning the magnetic damping in an active magnet via the magnetization of a passive magnetic layer, especially for insulating magnets. In addition to controlling the dynamics of the uniform mode, this magnetic `gate' concept~\cite{Cornelissen2018} can further be employed for modulating the magnon-mediated spin transport in a magnetic insulator~\cite{Cornelissen2015,Goennenwein2015}. The anisotropy in the resulting Gilbert damping may also offer a pathway towards dissipative squeezing~\cite{Kronwald2014} of magnetic modes, complementary to the internal anisotropy-mediated `reactive' squeezing~\cite{Kamra2016A,Kamra2017B}. We also found the longitudinal accumulated spin, which is often disregarded, to significantly affect the dynamics. This contribution is expected to play an important role in a wide range of other phenomena such as spin valve oscillators.


{\it Summary.} -- We have investigated the angular modulation of the magnetic damping in a free layer via control of the static magnetization in the fixed layer of a spin valve device. The damping can be engineered to become larger for either collinear or orthogonal configuration by choosing the longitudinal spin conductance of the fixed layer smaller or larger than its spin-mixing conductance, respectively. The control over damping is predicted to be sharp for spin valves made from insulating magnets. Our results pave the way for exploiting magneto-damping effects in spin valves.

{\it Acknowledgments.} -- We acknowledge financial support from the Research Council of Norway through its Centers of Excellence funding scheme, project 262633, ``QuSpin'', and from the Swedish Research Council, project 2018-03526, and Stiftelse Olle Engkvist Byggm{\"a}stare.

\bibliography{AnisoGB}


\widetext
\clearpage
\setcounter{equation}{0}
\setcounter{figure}{0}
\setcounter{table}{0}
\makeatletter
\renewcommand{\theequation}{S\arabic{equation}}

\begin{center}
	\textbf{\large Supplemental material with the manuscript Anisotropic and controllable Gilbert-Bloch dissipation in spin valves by} \\
	\vspace{0.3cm}
	Akashdeep Kamra, Dmytro M. Polishchuk, Vladislav Korenivski and Arne Brataas
	\vspace{0.2cm}
\end{center}

\setcounter{page}{1}


\section{Collinear configuration without longitudinal spin relaxation}
In order to appreciate some of the subtleties, we first examine the collinear configuration in the limit of vanishing longitudinal spin conductance. $\theta = 0,\pi$ and $g_l = 0$ imply the following values for the various parameters:
\begin{align}
	g_{xx} = g_{yy} = g_r^\prime, \quad g_{zz} = g_{yz} = p = 0, \quad l_{x,y} = \frac{g_r^\prime}{g_r + g_r^\prime} \equiv l,
\end{align} 
whence we obtain:
\begin{align}
	\frac{\mu_{sz}}{\hbar} = & \frac{ \left(m_x \dot{m}_y - m_y \dot{m}_x\right)}{m_x^2 + m_y^2}, \\
	= & \frac{\omega_0 + \omega_{ax}}{1 + \frac{\omega_{ax}}{2 \omega_0} \left[ 1 - \cos (2 \omega t)\right]},
\end{align}
where we have assumed magnetization dynamics as given by the Landau-Lifshitz equation without damping, and the phase of $m_x$ is treated as the reference and set to zero. In order to obtain analytic expressions, we make the assumption $\omega_{ax}/\omega_0 \ll 1$ such that we have:
\begin{align}
	\mu_{sz} = & \mu_{sz0} + \mu_{sz2}, \quad \quad \mathrm{with} \\
	\mu_{sz0} = & \hbar \left(\omega_0 + \frac{\omega_{ax}}{2}\right), \label{musz0} \\
	\mu_{sz2} = & \frac{\hbar \omega_{ax}}{4} \left( e^{-i 2 \omega t} + e^{i 2 \omega t} \right).
\end{align}
In contrast with our assumptions in the main text, a term oscillating with $2\omega$ appears. Furthermore, it yields contributions to the Bloch damping via products such as $m_y \mu_{sz}$, which now have contributions oscillating at $\omega$ due to the $\mu_{sz0}$ as well as $\mu_{sz2}$. We obtain:
\begin{align}
	\tilde{\alpha}_{xx} = \tilde{\alpha}_{yy} = & l, \\
	\tilde{\Omega}_{xx} = - l ~ \frac{\omega_0 + \frac{3 \omega_{ax}}{4}}{\omega_0 + \frac{\omega_{ax}}{2}} \quad \mathrm{and} &  \quad \tilde{\Omega}_{yy} = - l ~ \frac{\omega_0 + \frac{ \omega_{ax}}{4}}{\omega_0 + \frac{\omega_{ax}}{2}},
\end{align}
substituting which into Eq. (3) from the main text yields a vanishing linewidth and damping. This is expected from the general spin conservation argument that there can be no damping in the system if it is not able to dissipate the z-component of the spin. In fact, in the above considerations, $\mu_{sz2}$ contributed with the opposite sign to $\tilde{\Omega}_{xx}$ and $\tilde{\Omega}_{yy}$, and thus dropped out of the linewidth altogether. This also justifies our ignoring this contribution in the main text. 

\begin{figure}[tb]
	\begin{center}
		\includegraphics[width=65mm]{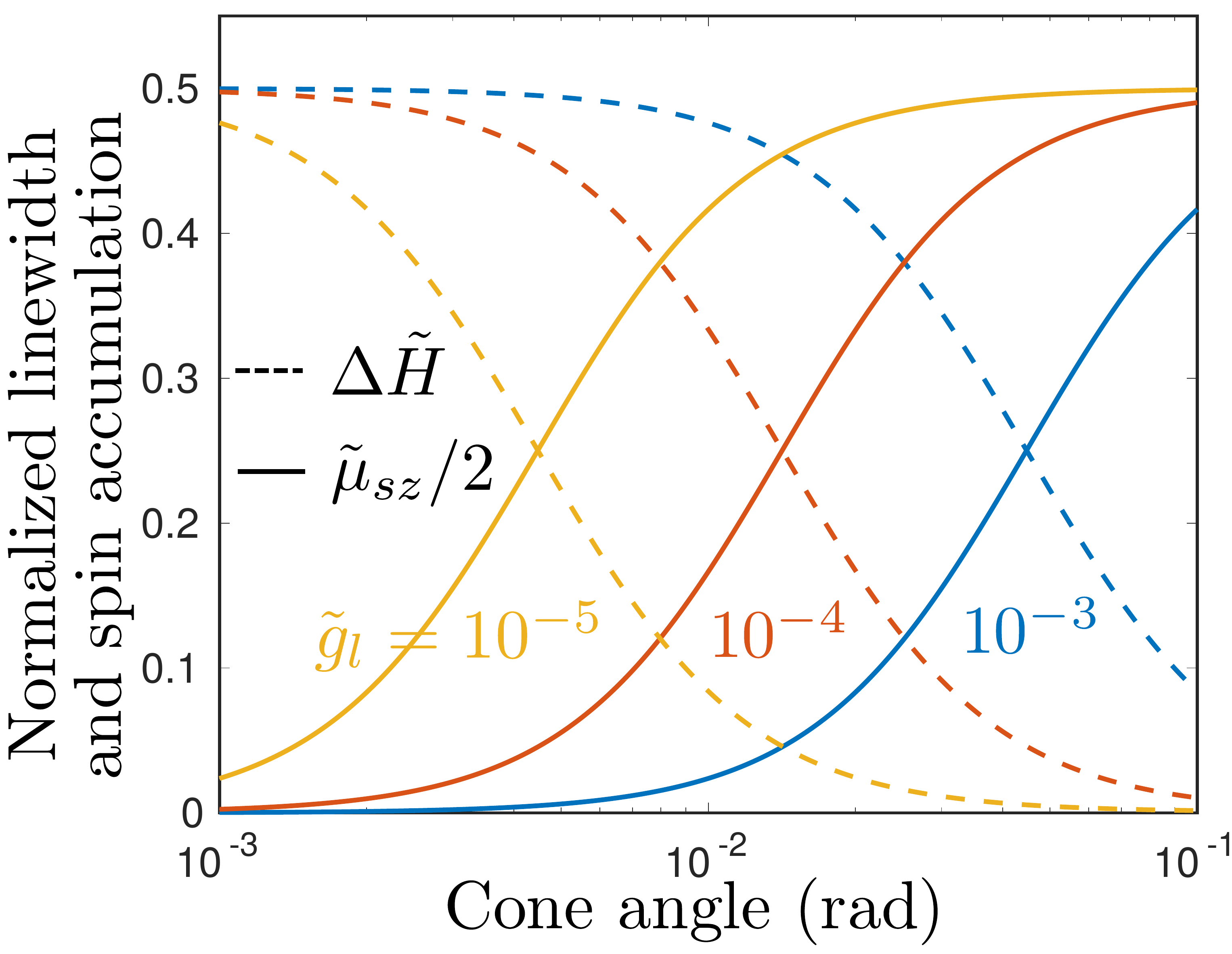}
		\caption{Ferromagnetic resonance linewidth and the dc spin accumulation created in the spacer as a function of the average cone angle in the collinear configuration. Depending on $\tilde{g}_l$, there is a complementary transition of the two quantities between small and large values as the cone angle increases. $\tilde{g}_r^\prime = 1$, $\omega_0 = 10 \times 2 \pi$ GHz, and $\omega_{ax} = 1 \times 2 \pi$ GHz.}
		\label{fig:coll}
	\end{center}
\end{figure}

Figure \ref{fig:coll} depicts the dependence of the accumulated z-polarized spin and the normalized linewidth for small but finite $g_l$ in the collinear configuration. The accumulated longitudinal (z-polarized) spin increases with the cone angle and the linewidth accordingly decreases to zero~\cite{Berger2001}.


\section{Numerical evaluation}

Despite the additional complexity in the previous section, we could treat the dynamics within our linearized framework. However, in the general case, $\mu_{sz}$ has contributions at all multiples of $\omega$ and cannot be evaluated in a simple manner. A general non-linear analysis must be employed which entails treating the magnetization dynamics numerically altogether. Such an approach prevents us from any analytic description of the system, buries the underlying physics, and is thus undesirable.  

Fortunately, the effects of non-linear terms are small for all, but a narrow, range of parameters. Hence, we make some simplifying assumptions here and continue treating our system within the linearized theory. We only show results in the parameter range where our linear analysis is adequate. Below, we describe the numerical routine for evaluating the various quantities. To be begin with the average cone angle $\Theta$ is defined as:
\begin{align}
	\Theta^2 = & \left\langle m_x^2 + m_{y}^2 \right\rangle,
\end{align} 
where $\langle \cdot \rangle$ denotes averaging over time. The spin accumulation is expressed as $\mu_{sz} = \mu_{sz0} + \mu_{sz1}$ with:
\begin{align}
	\mu_{sz0} = & \left\langle \frac{\hbar g_r \left( l_x m_x \dot{m}_y - l_y m_y \dot{m}_x - p \dot{m}_x \right)}{g_{zz} - p g_{yz} + g_r \left( l_x m_x^2 + l_y m_y^2 + 2 p m_y\right)} \right\rangle, \\
	\mu_{sz1} = & - \left\langle \frac{ g_r  p  }{g_{zz} - p g_{yz} + g_r \left( l_x m_x^2 + l_y m_y^2 + 2 p m_y\right)} \right\rangle ~ \hbar \dot{m}_x.
\end{align}
The above expressions combined with the equations for the spin current flow (Eqs. (6) in the main text) directly yield the Gilbert and Bloch damping tensors.


\section{Variation with additional parameters}

\begin{figure}[bt]
	\begin{center}
		\subfloat[]{\includegraphics[width=65mm]{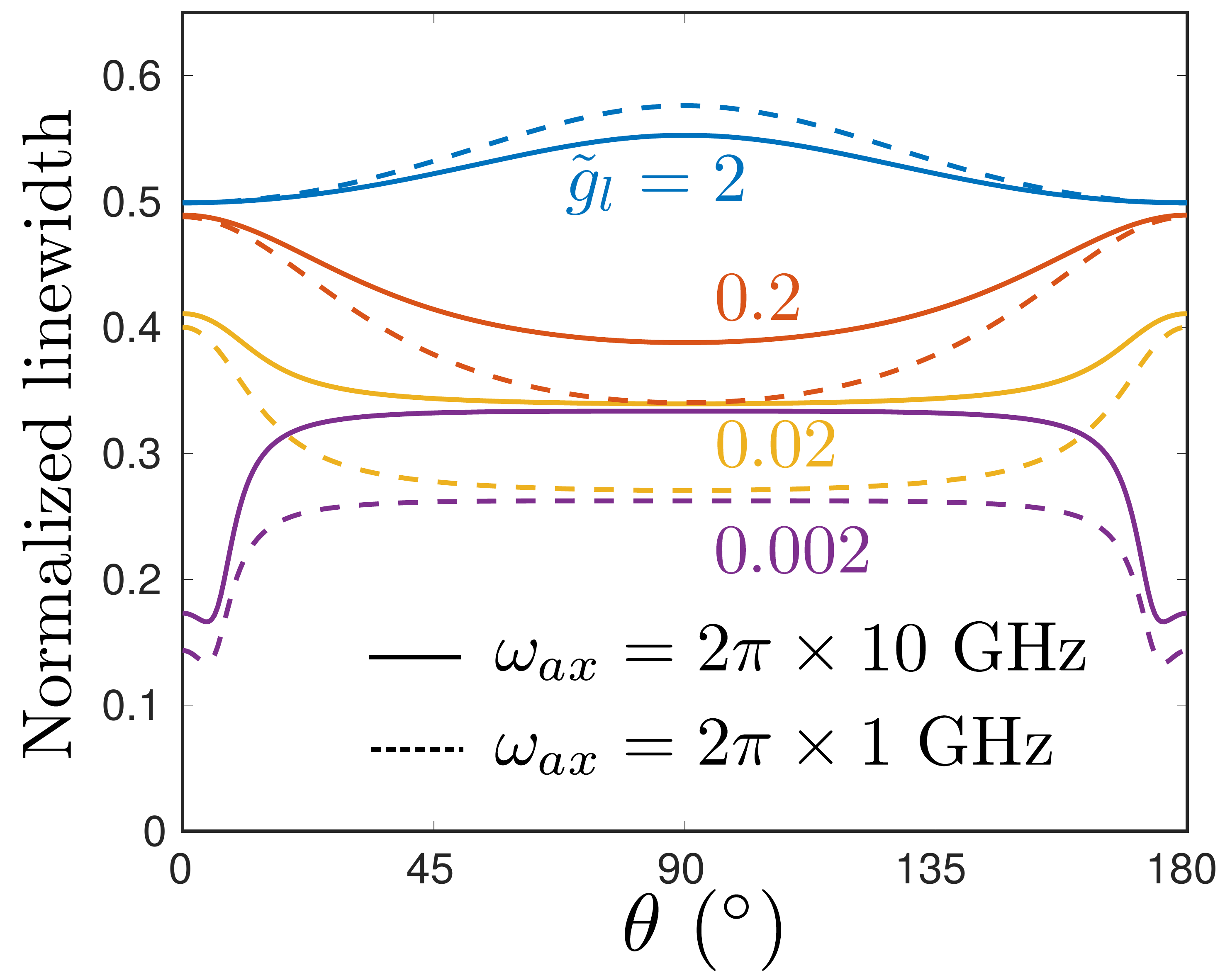}} \quad \quad
		\subfloat[]{\includegraphics[width=65mm]{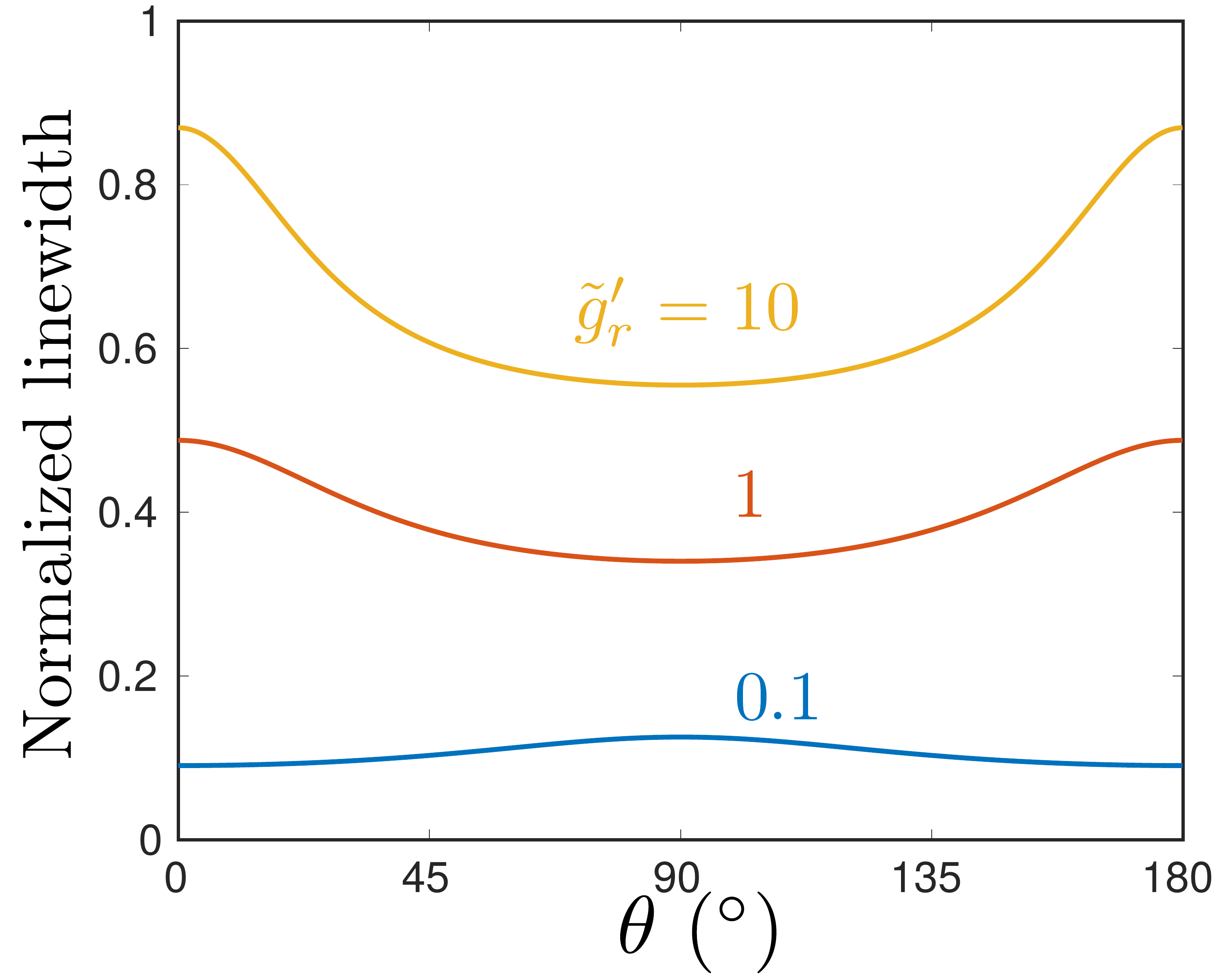}}
		\caption{Normalized ferromagnetic resonance (FMR) linewidth of $\mathrm{F}_1$. (a) Same as Fig. 3 in the main text with additional plots for a large easy-plane anisotropy. (b) Linewidth dependence for different spin-mixing conductances of N$|\mathrm{F}_2$ interface. The parameters employed are the same as Fig. 2 in the main text.}
		\label{fig:add_dep}
	\end{center}
\end{figure}

Here, we discuss the dependence of the FMR linewidth on the easy-plane anisotropy and the spin-mixing conductance $g_r^\prime$ of the N$|\mathrm{F}_2$ interface. The results are plotted in Fig. \ref{fig:add_dep}. A high easy-plane anisotropy is seen to diminish the configuration dependence of the linewidth and is thus detrimental to the dissipation tunability. The easy-axis anisotropy, on the other hand, is absorbed in $\omega_0$ and does not need to be examined separately. We also see an increase in the configuration dependence of the damping with an increasing $g_r^\prime$. This is understood simply as an increased damping when the spin is absorbed more efficiently due to a larger $g_r^\prime$. The damping is expected to reach the case of spin pumping into a perfect spin sink in the limit of $\tilde{g}_r^\prime \to \infty$ and $\theta = 0,\pi$.

\section{Effect of spin relaxation in the spacer layer}

\begin{figure}[tb]
	\begin{center}
		\includegraphics[width=50mm]{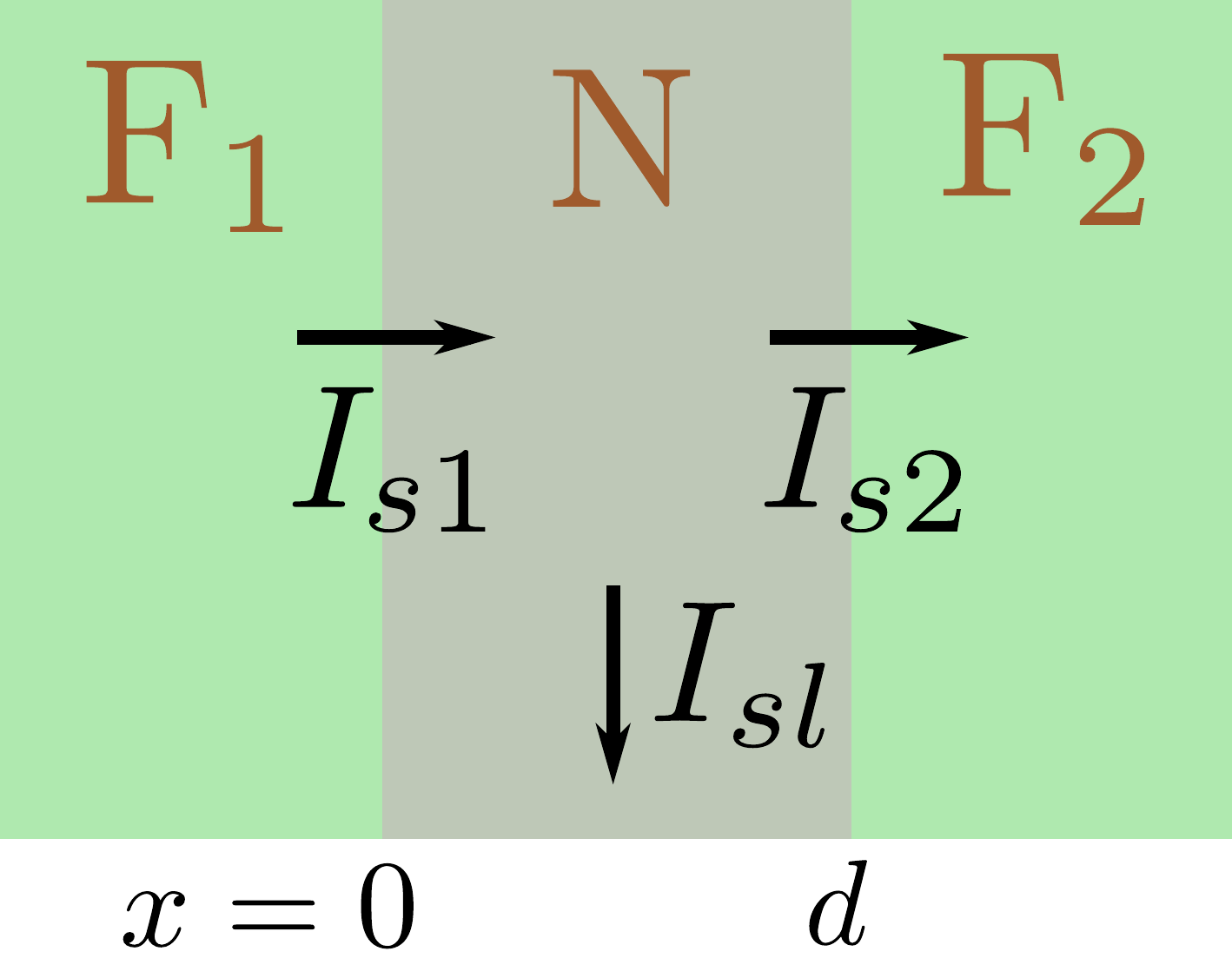}
		\caption{Schematic depiction of the spin currents flowing through the device, including the spin-leakage current $\pmb{I}_{sl}$ that is lost on account of a finite spin relaxation in the spacer layer N.}
		\label{fig:schematic_leakage}
	\end{center}
\end{figure}

We now address the role of the small but finite spin relaxation in the non-magnetic spacer layer. To this end, we consider that a part of the spin current injected into N by $\mathrm{F}_1$ is lost as the ``spin-leakage current'' $\pmb{I}_{sl}$, as depicted in Fig. \ref{fig:schematic_leakage}, such that $\pmb{I}_{s1} = \pmb{I}_{s2} + \pmb{I}_{sl}$. In order to evaluate the leakage, we consider the spin diffusion equation in N which reads~\cite{Tserkovnyak2005}:
\begin{align}
	D \partial_x^2 \pmb{\mu}_s = & \frac{\pmb{\mu}_s}{\tau_{sf}},
\end{align}
where $D$ and $\tau_{sf}$ are diffusion constant and spin-flip time, respectively. We now integrate the equation over the thickness of N:
\begin{align}
	\int d \left(D \partial_x \pmb{\mu}_s\right) = & \int_{0}^{d} \frac{\pmb{\mu}_s}{\tau_{sf}} dx.
\end{align}
Since the N-layer thickness $d$ is typically much smaller than the spin diffusion length in N (e.g., a few nm versus a few hundred nm for Cu), we treat $\pmb{\mu}_s$ on the right hand side as a constant. Furthermore, in simplifying the left hand side, we invoke the expression for the spin current~\cite{Tserkovnyak2005}: $\pmb{I}_{s} = (- \hbar \mathcal{N} S D / 2 ) \partial_x \pmb{\mu}_s$, with $\mathcal{N}$ the one-spin density of states per unit volume and $S$ the interfacial area. Thus, we obtain
\begin{align}
	\frac{2}{\hbar \mathcal{N} S} \left( \pmb{I}_{s1} - \pmb{I}_{s2} \right) = &  \frac{d}{\tau_{sf}} \pmb{\mu}_s,
\end{align}
which simplifies to the desired relation $\pmb{I}_{s1} = \pmb{I}_{s2} + \pmb{I}_{sl}$ with
\begin{align}\label{eq:sl}
	\pmb{I}_{sl} = & \frac{\hbar \mathcal{N} V_{\mathrm{N}}}{2 \tau_{sf}} \pmb{\mu}_s \equiv \frac{g_{sl}}{4 \pi} \pmb{\mu}_s,
\end{align}
where $V_{\mathrm{N}}$ is the volume of the spacer layer N. 

It is easy to see that accounting for spin leakage, as derived in Eq. (\ref{eq:sl}), results in the following replacements to Eqs. (6) of the main text:
\begin{align}
	g_{xx} \to g_{xx} + g_{sl}, \quad g_{yy} \to g_{yy} + g_{sl}, \quad g_{zz} \to g_{zz} + g_{sl}.
\end{align}
Since all our specific results are based on Eqs. (6) of the main text, this completes our assessment of the role played by spin relaxation in N. Physically, this new result means that the condition for no spin relaxation in the system, which was previously treated as $g_{l} \to 0$, is now amended to $g_l + g_{sl} \to 0$. This, however, does not affect the generality and significance of the key results presented in the main text.

\end{document}